\documentclass[%
 superscriptaddress,
 nofootinbib,
 amsmath,amssymb,
 aps,
 prx,
 twocolumn,
 floatfix,
]{revtex4-1}
\usepackage[normalem]{ulem}
\usepackage{graphicx}
\usepackage{dcolumn}
\usepackage{bm}
\usepackage{physics}
\usepackage{xfrac}
\usepackage{bbm, ulem}
\usepackage{hyperref}

\usepackage{titlesec}
\titleformat{\paragraph}[runin]
        {\bfseries}
        {}
        {0.0em}
        {}
        [ -- ~]
\titlespacing*{\paragraph}{0pt}{4pt}{0pt}

\usepackage{color}
\usepackage[dvipsnames]{xcolor}

\begin{document}

\preprint{APS/123-QED}
\title{Holographic quantum simulation of entanglement renormalization circuits}

\author{Sajant Anand}
\email{sajant@berkeley.edu}
\affiliation{%
Department of Physics, University of California, Berkeley, CA 94720, USA
}%

\author{Johannes Hauschild}
\affiliation{%
Department of Physics, University of California, Berkeley, CA 94720, USA
}%

\author{Yuxuan Zhang}
\affiliation{%
Department of Physics, University of Texas at Austin, Austin, TX 78712, USA
}%

\author{Andrew C. Potter}
\affiliation{Department of Physics and Astronomy, and Stewart Blusson Quantum Matter Institute,
University of British Columbia, Vancouver, BC, Canada V6T 1Z1}

\author{Michael P. Zaletel}
\email{zaletel@berkeley.edu}
\affiliation{%
Department of Physics, University of California, Berkeley, CA 94720, USA
}%
\affiliation{%
Material Science Division, Lawrence Berkeley National Laboratory, Berkeley, CA 94720, USA
}%

\date{\today}

\begin{abstract}

While standard approaches to quantum simulation require a number of qubits proportional to the number of simulated particles, current noisy quantum computers are limited to tens of qubits.
With the technique of holographic quantum simulation, a $D$-dimensional system can be simulated with a $D{\rm -}1$-dimensional subset of qubits, enabling the study of systems significantly larger than current quantum computers.
Using circuits derived from the multiscale entanglement renormalization ansatz (MERA), we accurately prepare the ground state of an $L=32$ critical, non-integrable perturbed Ising model and measure long-range correlations on the 10 qubit Quantinuum trapped ion computer. 
We introduce generalized MERA (gMERA) networks that interpolate between MERA and matrix product state networks and demonstrate that gMERA can capture far longer correlations than a MERA with the same number of qubits, at the expense of greater circuit depth. 
Finally, we perform noisy simulations of these two network ans\"atze and find that the optimal choice of network depends on noise level, available qubits, and the state to be represented.
\end{abstract}

\maketitle

\section{Introduction}
\label{sec:introduction}
Simulations of strongly-correlated quantum many-body physics relevant to materials science, chemistry, and fundamental physics are promising early applications of quantum computers~\cite{preskill2018}. 
An important task is to approximately prepare the ground state of a given system and measure its properties. 
Ground-state preparation is also a prerequisite for exploring near-equilibrium quantum dynamics of transport and AC responses.
Standard approaches to quantum simulation directly encode each spin- or electron-orbital into a distinct hardware qubit, restricting the accessible problem-size to the available qubit number.
The paradigm of \textit{holographic quantum simulation} provides an alternative for simulating a system of $n$ sites with less than $n$ qubits.
Here, selected qubits are regularly measured, reset, and reused throughout the computation, and correlations between different physical sites are mediated by unmeasured qubits~ \cite{schon2005sequential, kim2017, kim2017b, barratt2020parallel, Foss-Feig2021a}. 
This allows a $D$-dimensional system to be simulated on a quantum computer with only enough hardware qubits to store the state of a $(D{\rm-}1)$-dimensional cross-section\footnote{This notion of holography is distinct from the notion of holographic field-theory/gravity correspondences.}. 
Formally, these techniques can be understood as a quantum compression of states into matrix-product state (MPS) or tensor network state (TNS) form ~\cite{schon2005sequential,barratt2020parallel, Foss-Feig2021a}.

MPS \cite{ostlund1995} provide an efficient representation of 1D states with area-law entanglement, with an accuracy that improves with the rank of the tensors $\chi$ faster than any power-law \cite{verstraete2006, hastings2007}.
The classical cost of MPS algorithms scale as $\mathcal{O}(\chi^2)$ for storage and $\mathcal{O}(\chi^3)$ for time, with $\chi$ growing exponentially with the bipartite entanglement entropy $S$.
While tractable for most gapped 1D ground states, when simulating quantum dynamics or strips of $D>1$-dimensional models, $S$ grows linearly in time or cross-sectional area, quickly making classical simulations  prohibitive. 
The  holographic approach for quantum simulation of an MPS requires only $1 + \log_2(\chi) \sim \mathcal{O}(S)$ qubits, an exponential reduction in storage. The time cost, however, is more subtle to assess, since holographically preparing a \emph{generic} MPS requires acting with a $\chi \times \chi$ unitary which  must be broken into a very deep circuit of gates. 
This obstacle motivated recent work \cite{zaletel2020,haghshenas2021,lin2021,slattery2021,wei2022} to investigate the variational power of MPS and 2D TNS with tensors generated by finite-depth quantum circuits, dubbed quantum MPS (qMPS) and quantum TNS (qTNS), respectively. It has been found that such circuits provide an efficient parameterization for approximating interesting physical states, sometimes even representing states with fewer variational parameters than their generic (dense) tensor network counterpart without sacrificing accuracy.
Additionally for 1D systems, ground state optimization for correlated spin~\cite{Foss-Feig2021a} and electron~\cite{niu2021} systems, time evolution~\cite{chertkov2021,lin2021}, and many-body entanglement measurements~\cite{Foss-Feig2021b} were demonstrated experimentally on trapped-ion and superconducting qubit devices using holographic simulation of qMPS. 

While finite-$\chi$ MPS can faithfully represent ground states of thermodynamically-large 1D gapped systems~\cite{hastings2007}, the natural tensor network for 1D critical systems is the multiscale entanglement renormalization ansatz (MERA), which can  capture the power-law decay of critical correlations and the critical scaling of the entanglement entropy $S(L) \sim \log(L)$ of a subsystem of size $L$ ~\cite{vidal2007,vidal2008,evenbly2013,evenbly2014}. 
While MPS and MERAs for 1D critical states both require a number of qubits which grows logarithmically with the system size to account for the growth in entanglement,  MERAs are exponentially more sparsely parameterized than MPS, due to the inherent tree structure and thus avoid the blow-up in quantum gate count.

In this work, we holographically prepare a generalization of the MERA representing critical, non-integrable ground states on Quantinuum's System Model H1 trapped-ion quantum computer~\cite{pino2021}.
We find that increasing the MERA depth leads to measurable improvements in the observed long-range correlations. 
We discuss an enlarged class of generalized MERA (gMERA) which provide a natural interpolation between MERA and MPS networks.
We holographically prepare the gMERA and demonstrate that quantum correlations can be accurately measured, even using a low-depth network. 
Finally, we discuss the merits of these two networks in the presence of varying levels of noise.

\section{Holographic simulation of (g)MERA}
\label{sec:method}
There are two routes for holographically simulating a MERA. 
The first is to prepare a local patch of the desired state in a ``top-down" approach~\cite{kim2017c}. 
As any local subset of sites in a MERA exhibit a bounded past causal cone, i.e. the set of tensors at each scale that can affect the subset, one can prepare the subset's reduced density matrix by viewing the MERA as a fine-graining quantum channel. 
Starting from fixed input states, one repeatedly applies the gates of the MERA, discarding and reusing those that exit the bounded causal cone of the desired patch. 
After sufficient iterations, the state of the qubits will approach the fixed point of the quantum channel, even with noisy gate operations. 
This approach was recently demonstrated on a trapped ion quantum computer \cite{sewell2021}, finding that such circuits are robust to noise as theoretically predicted in~\cite{kim2017c}.

The second method is to view the MERA ``sideways" as a circuit for preparing each site in turn, effectively trading the space and time directions.
As in holographic simulations of qMPS, bond qubits carry correlations between sites represented by a physical qubit.
Analogous theoretical guarantees of noise robustness as in the top-down approach cannot be made, as the circuit is not used to prepare fixed points of the quantum channel.
As discussed later, however, we find such circuits to be remarkably resilient to noise.

\begin{figure}
    \centering
    \includegraphics[width=0.47\textwidth]{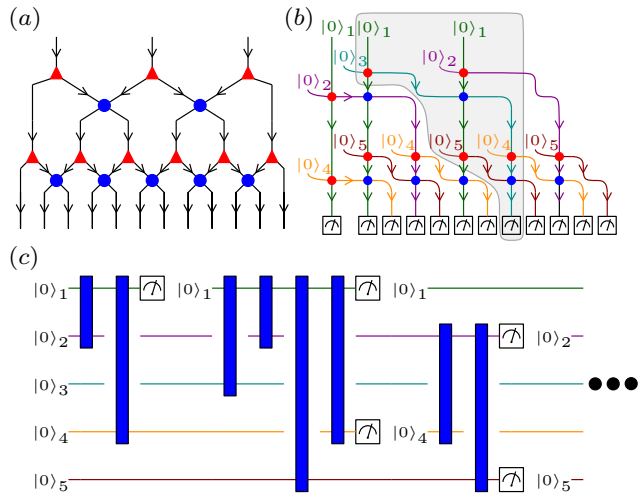}
    \caption{Holographic simulation of a MERA. \textbf{(a)} Depth 2 MERA network structure for a spin chain of 8 sites consisting of alternating sublayers of two-site unitaries (blue) and isometries (red). \textbf{(b)} Isometric representation for the MERA formed by bending the wires. Five qubits, which are color coded, are required for holographic simulation. Arrows indicate the flow of time in the quantum circuit. The causal cone for site $8$ is shaded in grey, and the number of gates/tensors in the causal cone is independent of the length $L$ of the state. \textbf{(c)} Quantum circuit for the first five sites corresponding to 
    the MERA network. Each isometry has been enlarged to a two-qubit unitary, and mid-circuit measurements and resets are shown explicitly.
    }
    \label{fig:MERA}
\end{figure}

We focus on binary MERAs as shown in Fig.~\ref{fig:MERA}(a), where each layer consists of both a row of disentangling unitaries and coarse-graining isometries which halve the degrees of freedom \cite{vidal2008, evenbly2009, evenbly2013}.
While MERAs are often used in their scale-invariant formulation with a single layer repeated indefinitely, the number of available qubits limits a sideways holographic simulation to MERAs with a finite depth of $d$ layers. 
Such networks have a finite support for correlations, growing exponentially with $d$; the maximum range $r_d$ of nonzero correlations in the bulk is given by the recursion relation 
\begin{align}
    r_d = 2 r_{d-1} + 2 \text{ with } r_0 = 1.
    \label{eq:recursion}
\end{align} 
Note that while doubling the range of correlations, each additional layer in the MERA requires only roughly half the number of tensors as the previous, leading to a total number of $\sum_{j=1}^d \left( L/2^{j-1} - 1 \right) < 2L$ tensors for a lattice of $L=2^n$ sites.

\subsection{Quantum circuit implementation}
Given an optimized binary MERA representing a state that we wish to holographically prepare, we first reshape the network to the \textit{isometric} form shown in Fig.~\ref{fig:MERA}(b).
This is accomplished simply by bending the legs of the diagram so that the isometric arrows point down and to the right and is done to be consistent with the generalized MERA (gMERA) networks we discuss later.
To prepare the state on a quantum computer, we extend the isometries with orthogonal columns into square unitary matrices generated by gates acting on a physical qubit and ancilla bond qubits. 
When extending an isometry, the added legs act on a fixed reference state, $\ket{0}$, as shown in Fig.~\ref{fig:MERA}(b).
The circuit propagates along the direction of the arrows~\footnote{This is opposite to the convention in tensor network literature where incoming arrows indicate isometry conditions on the tensor.}, so the leftmost site is prepared first starting with $\ket{0}$ initialized qubits and acting with the gates specified by the tensors moving down the leftmost column.
Once all tensors in the column have been used, we can measure the physical qubit in the desired basis and proceed to the next site.
Once measured, the physical qubit can be reset to $\ket{0}$ and re-used in other parts of the circuit, as depicted in Fig.~\ref{fig:MERA}(c).

For a bond dimension $\chi$ MERA, each gate acts on $2\log_2{\chi}$ qubits, so we require $\chi=2$ to work with two-site gates.
Hence a depth $d$, $\chi=2$ MERA circuit requries $2d+1$ qubits, which can be intuitively understood as 1 qubit for each isometric and unitary sublayer and 1 physical qubit. 
For implementation on quantum computers with a limited two-qubit gate set, we decompose each $SU(2)$ gate into a sequence of 7 single-qubit rotations specified by 15 angles and three CNOTs, ignoring an irrelevant global phase~\cite{vatan2004}. 
Note that in the interior of the MERA, one can exploit the gauge freedom on contracted indices of the MERA to reduce this further to just 9 free angles per two-qubit unitary. 
For models with charge conservation, the number of free parameters can be reduced even further by enforcing a block-diagonal form of the tensors, similarly to as in classical tensor network simulations \cite{singh2010,niu2021}.

The causal cone of a MERA has bounded width, as shown in Fig.~\ref{fig:MERA}(b).
The causal cone for binary MERA saturates to width 3 after a few coarse-graining layers, so 6 qubits are sufficient to evaluate up to next-nearest neighbor operators for the top-down holographic method regardless of circuit depth $d$.
Generic $n$-point functions will require $6n$ qubits, as $n$ disjoint causal cones must be maintained until they merge.
The sideways holographic approach uses a fixed $2d + 1$ qubits to evaluate arbitrary operators, as each site is prepared sequentially.
Hence evaluation of correlation functions as a function of distance between operators is more naturally implemented on qubit-limited devices using the sideways methods.

As holographic simulation enables preparation of the parameterized tensor network on the quantum hardware and measurement of all desired expectation values, one can imagine an optimization scheme where one iteratively adjusts parameters of the network and measures the energy as cost function to be minimized directly on the quantum hardware \cite{kim2017,Moll2018,slattery2021}.
The sideways approach to holographic MERA simulation lends itself to such optimization, as for the example of the transverse-field Ising model discussed later, two iterations of the circuit, measuring once in the $X$ and once in the $Z$ basis, are sufficient to determine the energy and all N-point $X$ and $Z$ correlations, even in spatially inhomogeneous systems.
An iteration of a sideways MERA circuit requires at most $2L$ gates for a maximum depth circuit. 
For the top-down method, preparing each site requires approximately $5d$ gates, yielding a total gate cost of $5dL$.
To reduce the quantum hardware cost, in this work we instead optimize the tensor networks with classical computers; see Appendix~\ref{sec:optimization} for details on how we determine the classical tensor networks and convert the tensors into two-qubit gates.

\begin{figure}
    \centering
    \includegraphics[width=0.47\textwidth]{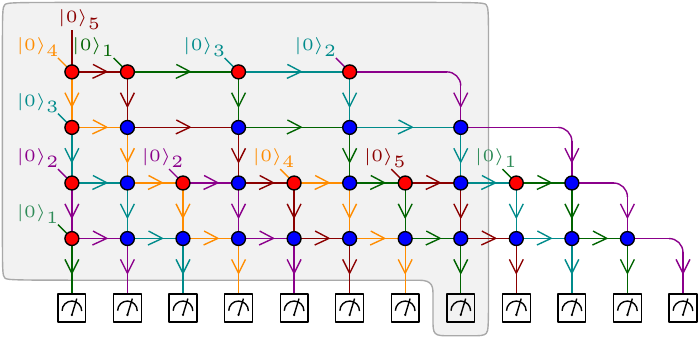}
    \caption{Isometric representation for gMERA structure of depth 2 for a chain of 12 sites. Five qubits, which are color coded, are required for holographic simulation. The red and blue tensors are isometries and unitaries, respectively. The causal cone for site $8$ is shaded in grey, and for a state of length $L$, there are $\mathcal{O}(L)$ gates/tensors in the causal cone, regardless of depth.}
    \label{fig:isoMERA}
\end{figure}

\begin{figure*}
    \centering
    \includegraphics[width=\textwidth]{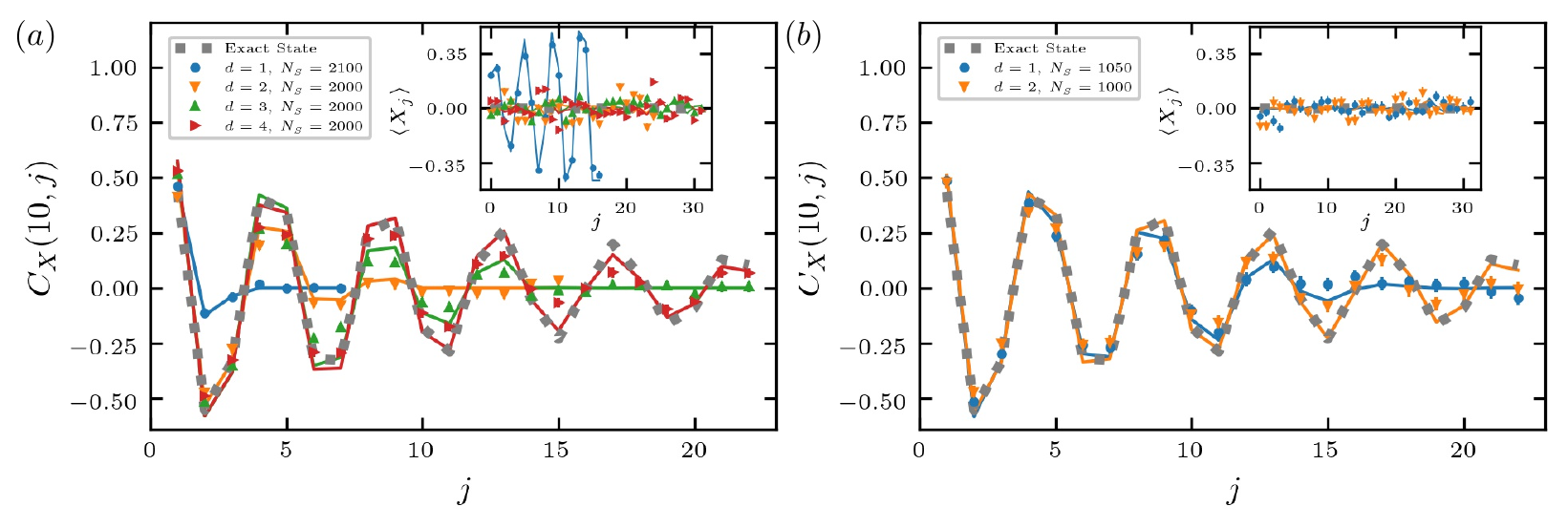}
    \caption{Connected $XX$ correlations $C_{X}(10, j)$ for a $L=32$ $H_{\text{TFI-SD}}$ chain. In both plots, solid lines are the exact tensor network results while dots with error bars are from $N_S$ measurement shots on the Quantinuum H1 computer for \textbf{(a)} MERA and \textbf{(b)} gMERA networks of varying depth. Insets show $\langle X \rangle$ for the same networks. We simulate up to site $17$ for the depth 1 MERA, site $25$ for depth 2, and the entire chain for depths 3 and 4 MERA and depths 1 and 2 gMERA.
    } \label{fig:TFISD}
\end{figure*}

\subsection{Generalized MERA (gMERA)}
The finite support of correlations inherent in a MERA motivates us to consider an alternate ansatz which we call the generalized MERA (gMERA). As illustrated in Fig.~\ref{fig:isoMERA}, each layer of this network can be viewed as a coarse-graining matrix product operator, first introduced in the context of MPS renormalization \cite{bal2016}.
The gMERA interpolates between the MERA and circuits composed of ladder unitary layers that form a simple implementation of qMPS (we consider such ladder circuits in Appendix~\ref{sec:ladder}).
Any MERA with bond-dimension $\chi$ can also be represented by a gMERA with the same vertical bond dimension by simply replacing selected gates in \ref{fig:isoMERA} with identity operations to produce the circuit shown in Fig.~\ref{fig:MERA}(b).
The gMERA corresponds to switching from a brick-wall pattern of unitaries used in the MERA to a ladder pattern, so already a single layer of unitaries has an infinite support of correlations \cite{lin2021}.
Yet, correlations of a single layer still decay exponentially as for MPS \cite{bal2016}.
Further, we note that even $\chi > 2$ binary MERA can be represented by a gMERA with vertical bond dimension 2 but horizontal bond dimension $\chi^2 / 2$. 

Like the regular MERA, a $\chi=2$ gMERA with $d$ layers requires  $2d + 1$ qubits in a holographic simulation, 
which again can be performed column by column as before.
For a lattice of $L=2^n$ sites, a depth $d$ gMERA requires $\sum_{j=1}^d \left( L/2^{j-2} - 3 \right) < 4L$ arbitrary two-qubit gates, roughly twice as many as the MERA. 
This additional gate cost is contrasted with the infinite support of correlations even with a single gMERA layer.
As can be seen in Fig.~\ref{fig:isoMERA}, the causal cone for any site includes all tensors up and to the left and thus is extensive.
Top-down simulations are not feasible for gMERA circuits, as the required number of qubits depends on both $d$ and $L$.

We note that we could instead consider (g)MERA networks with $\chi>2$ and instead parameterize each tensor in terms of two-qubit gates~\cite{haghshenas2021}. A recent study found qMPS tensors parameterized in this way to, at times, provide a more efficient while equally accurate representation of the ground state than dense MPS tensors of dimension $\chi \times D \times \chi$, where $D$ is the physical Hilbert space dimension.

\begin{figure*}
    \centering
    \includegraphics[width=\textwidth]{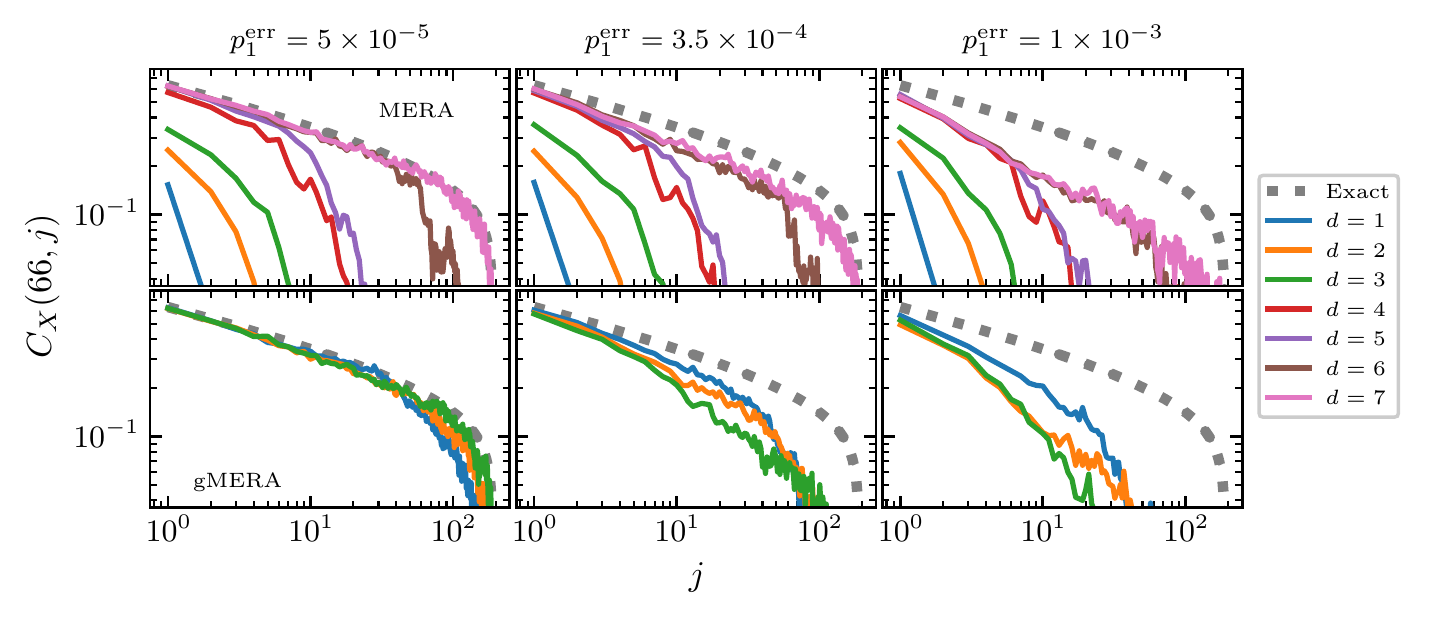}
    \caption{Simulations of ground states for the unperturbed TFI model realized by MERA (top row) and gMERA (bottom) networks for different probabilities of depolarizing noise. Two qubit error rates are chosen as $p^{\mathrm{err}}_{2} = 10 p^{\mathrm{err}}_{1}$. For both networks of depth $d$, the required number of qubits for holographic simulation is $2d + 1$, indicating that low depth gMERA can capture longer correlations than same depth MERA counterparts, even in the large noise limit.}
    \label{fig:noise}
\end{figure*}

\section{Hardware implementation}
\label{sec:results}
To demonstrate the ability of quantum circuits generated from (g)MERA to accurately capture quantum correlations using these holographic simulation methods, we study the spin-1/2 transverse field Ising (TFI) model with a \textit{self-dual} perturbation:
\begin{align}
    H =& \sum_{i} -\left(X_i X_{i+1} + Z_{i} \right) + V \left(Z_i Z_{i+1} + X_{i} X_{i+2} \right)
    \label{eq:tfisd}
\end{align}
where $X_i, Z_i$ are the spin-$\sfrac{1}{2}$ Pauli operators on site $i$. 
The first term is the critical TFI model which is  integrable.
To break integrability, we add a self-dual coupling, which under the Kramers-Wannier duality $X_i X_{i+1} \rightarrow \tilde{Z}_{i}, Z_i \rightarrow \tilde{X}_{i}\tilde{X}_{i+1}$ maps to itself \cite{radicevic2019spin}. 
Coupling $V \approx 250$ corresponds to the tri-critical point described by the tri-critical Ising CFT with central charge $c=7/10$. 
Values $-0.28 < V < 250$ correspond to critical, non-integrable perturbations to the critical Ising CFT with usual central charge $c=1/2$ \cite{rahmani2015}. 
In our experiments and simulations, we use $V = 4$ to maximize the magnitude of the oscillations of $XX$ correlation functions. 

Holographic simulation requires the ability to address, measure, and reset individual qubits throughout the quantum circuit and thus is naturally suited to trapped ion devices.
We run our circuits on the Quantinuum's System Model H1 quantum computer, with ten $^{171}\text{Yb}^+$ hyperfine qubits and a quantum volume of 64 \cite{pino2021}. 
Given that a depth $d$ (g)MERA requires $2d+1$ qubits, we can prepare at most a depth-$4$ network using nine qubits for a lattice of arbitrary size. 
Such depth is sufficient for a MERA network to have non-zero correlations over a range of 46 sites.

We prepared $L=32$ ground states of the self-dual Ising model in Eq.~\eqref{eq:tfisd} as depths $d=1,2,3,4$ MERA and depths $d=1,2$ gMERA on the H1 quantum processor and measured all sites in the $X$ basis, extracting both on-site expectation value $\langle X_i \rangle$ and connected two-point correlation $C_{X}(10, r) = \langle X_{10} X_{10+r} \rangle - \langle X_{10} \rangle \langle X_{10+r} \rangle$.
Here, $i=10$ is chosen sufficiently within the bulk of our $L=32$ system to minimize boundary effects and also to reach the maximal support of non-zero correlations for depth $d=1,2,3$ MERA. 
For each MERA (gMERA), we obtained at least $N_S=2000$ ($N_S=1000$) samples of measurement bitstrings, which yielded expectation values and correlation functions simultaneously.
Additional experimental details can be found in the Appendix.

The results are shown in Fig.~\ref{fig:TFISD}.
We observe that increasing the depth of the MERA clearly extends the range of correlations we can capture, as the finite correlation support at low depth $d$ is clearly a severe limitation.
In stark contrast, any gMERA will have non-zero, exponentially-decaying correlations between any two sites, and we find that even a depth $d=1$ gMERA can capture the correlations fairly accurately over a wide range of sites much larger than the range $r_1=4$ of the non-zero MERA correlations.
A depth $d$ MERA uses approximately as many gates as a depth $d/2$ gMERA and twice as many qubits, and for this short chain, we see that both the $d=4$ MERA and $d=2$ gMERA can sufficiently capture correlations over the length of the chain.

\section{Noise Analysis}
\label{sec:noise}
We have shown experimentally that additional layers in either the MERA or gMERA network allow us to more accurately capture correlations between distant sites, as the network ansatz becomes more expressive.
Additional layers, however, incur a cost of increased number of two-qubit arbitrary unitary gates that must be performed, which leads to a larger negative effect of gate noise.

To investigate the tradeoff between network depth and performance on noisy quantum computers, we perform noisy simulations of MERAs and gMERAs of a $L=256$ unperturbed TFI model (Eq.~\ref{eq:tfisd} with $V=0$)~\footnote{We choose this Hamiltonian as the $XX$ correlations monotonically decrease rather than oscillate and thus are easier to analyze.} ground state and measure $C_X(66, r)$, as shown in Fig.~\ref{fig:noise}.
We study MERAs of depth up to 7 -- reducing the sites in the lattice from 256 sites down to 2 -- and gMERAs up to depth 3, where additional depths only lead to marginal improvements in the correlations; additional results for unitary networks formed entirely of ladder unitary sublayers of gMERA are discussed in Appendix~\ref{sec:ladder}.
Noisy circuit simulations are performed with Qiskit~\cite{Qiskit}, using $10^4$ measurement shots per data point, and with depolarizing errors  $E(\rho_{n}) = (1 - p^{\mathrm{err}}_{n}) \rho + p^{\mathrm{err}}_{n} \mathbbm{1}/ 2^{n}$ applied to all one- and two-qubit gates with depolarizing rate $p^{\mathrm{err}}_{1,2}$ respectively. 
On the Quantinuum processor, the error rates measured by randomized benchmarking were $p^{\mathrm{err}}_{1} = 1.1 \times 10^{-4}$ and $p^{\mathrm{err}}_{2} = 7.9 \times 10^{-3}$ \cite{pino2021}.
In simulations, we vary the one qubit error rate and fix the ratio between one and two qubit error rates to: $p^{\mathrm{err}}_{2} \equiv 10\cdot p^{\mathrm{err}}_{1}$. 

From the simulations shown in Fig.~\ref{fig:noise}, we find that for the three noise levels shown, there is always an improvement in the measured correlations by increasing the depth of the MERA circuit. 
This is because the added benefit of a more expressive ansatz with a longer ranger of non-zero correlations outweighs the added noise of additional gates; especially between a $d=6$ and $d=7$ network, the additional layer allows us to accurately capture correlations across the length of the chain while only adding 3 additional gates.

For the gMERA, where already a depth $d=1$ network can accurately capture correlations up to $\approx$ 30 sites at large noise levels, we expect additional layers to be more impacted by noise than additional MERA layers.
A depth $d$ gMERA uses approximately twice as many arbitrary two-qubit unitaries as a depth $d$ MERA.
Additionally, noisy tensors in isometric networks can only affect the tensors in their future causal cone.
For the MERA, the extent of the future causal cone of each tensor is finite and is governed by the recursion relation given earlier; for the gMERA, the future causal cone of each tensor is extensive, and thus an error at site $i$ will affect all sites $j > i$. 
Such effects can be seen in Fig.~\ref{fig:noise}, where the performance of higher-depth gMERAs is strongly dependent on the noise level, unlike that of high-depth MERA. This is consistent with previous analysis of the top-down holographic MERA approach, which was shown both theoretically and experimentally to be noise-resilient \cite{kim2017c, sewell2021}.

For MERA, across all noise levels, one should choose the minimum depth network -- in order to minimize qubit and gate costs -- that reproduces the noiseless correlations on the desired scale and to the desired accuracy; empirically we see that even with high noise rates a depth $d$ MERA more accurately captures correlations across all scales than networks with fewer layers.
For gMERA, the error rate fixes an optimal depth that minimizes the correlation errors; for example in Fig.~\ref{fig:noise}, for correlations between sites separated by a distance of 100, at low noise level the depth 3 gMERA is most accurate, while at high noise levels the accumulated noise from extensive future causal cones leads to depth 1 gMERA being optimal.
We note, however, that at all noise levels the depth $d$ gMERA performs better than the MERA of the same depth and number of qubits, as the benefit of an infinite range of correlations far outweighs the increased susceptibility to noise.
When, as is typically the case, the simulations are limited by the number of available qubits, the gMERA is hence a better choice, even in the presence of significant noise.
We note that all of the investigated circuits are derived from tensor networks optimized without noise, so it is an open question if better performance can be achieved by optimizing a circuit for a particular $p_{\mathrm{err}}$.

\section{Conclusion}
In this work, we have accurately measured long-range correlations in the ground state of a non-integrable critical 1D system represented by MERA and gMERA networks.
We demonstrated that increasing the depth of the network directly improved the correlations measured. 
The gMERA network interpolates between MERA and MPS and can accurately capture long-range quantum correlations even with low depth networks. 
Through noisy simulations, we confirm that MERA networks are resilient to noise and typically benefit from additional layers, while additional layers in the gMERA may not provide a benefit depending on the noise level and desired correlations.

We conclude by commenting on the extension of these techniques to other tensor networks and future directions.
Any tensor network composed entirely of isometric tensors can be simulated holographically, including the recently developed isometric tensor networks in two-dimensions which  represent non-chiral gapped states.\cite{zaletel2020, soejima2020, slattery2021,wei2022sequential}. 
It is not known whether isometric projected entangled pair states (isoPEPS) can be contracted efficiently, so studying such systems via quantum computers may provide a natural route to quantum advantage \cite{schuch2007}.
One limitation of the work proposed here is that we optimize the tensor network classically first, and only then convert it into a quantum circuit. While optimizing variational holographic circuits in a hybrid classical-quantum manner has been proposed and explored~\cite{Foss-Feig2021a,slattery2021,peruzzo2014}, it is still an open question on how to do this efficiently, especially as the depth of the circuit increases and while varying the level of noise. 
An interesting approach would be to investigate the use of direct sampling methods, which have recently shown to be effective in classically optimizing PEPS, on a quantum computer for isometric tensor networks \cite{vieijra2021}.

\vspace{4pt}\noindent{\it Acknowledgments -- }
We thank Garnet Chan, Michael Foss-Feig, and David Hayes for insightful conversations, and the Quantinuum trapped-ion experimental team for their assistance implementing hardware demonstrations.
SA acknowledges support from the Department of Defense (DoD) through the National Defense Science \& Engineering Graduate (NDSEG) Fellowship Program .
AP acknowledges support from the US Department of Energy (DOE) grant DE-SC0022102, and the Alfred P. Sloan Foundation through a Sloan Research Fellowship. 
JH and MZ acknowledge support from the NSF OIA Convergence Accelerator Program under award number 2040549.
This research was undertaken thanks, in part, to funding from the Max Planck-UBC-UTokyo Center for Quantum Materials and the Canada First Research Excellence Fund, Quantum Materials and Future Technologies Program.

\appendix
\section{Optimization of (g)MERA}
\label{sec:optimization}
To prepare the states on the quantum computer, we need a quantum circuit that can be run holographically. We first classically optimize the MERA and gMERA tensor networks. Instead of optimizing each network to minimize the energy, we optimize the networks to have maximal overlap with a reference ground state in the form of a high bond-dimension MPS.
This MPS is found by the density matrix renormalization group (DMRG) algorithm implemented in the TeNPy package \cite{white1992,hauschild2018}. A bond dimension $\chi=256$ is enough to represent the MPS to a very high accuracy for the small systems of $L=32$ sites.
We then subsequently build the (g)MERA networks by splitting off unitaries and isometries layer by layer as explained below, reducing the remaining physical sites in the MPS by a factor of 2 with each layer. Finally, we optimize this initial guess for the (g)MERA networks further by iterating over the individual tensors.

\subsection{Splitting off a MERA layer}
To obtain one layer of the MERA from the MPS of length $L'$, we first perform one sweep left to right, where we split off unitaries acting on every pair of two neighboring sites  $2n, 2n+1$, indexing the sites starting at $1$. These unitaries are chosen to disentangle the second Renyi entropy for a cut between the two sites \cite{Hauschild2018b}.
The found unitaries are used to disentangle the MPS, while their hermitian conjugates form the unitaries in the layer of the MERA.

Afterwards, we sweep right to left, splitting off the isometries on the other pairs of neighboring sites $2n-1, 2n$. We form the two-site orthogonality center $\Theta^{\sigma_{2n-1}, \sigma_{2n}}_{\alpha_{2n-1}, \alpha_{2n+1}}$, indexed by the local basis states $\sigma_{2n-1}, \sigma_{2n}$ on sites $2n-1, 2n$ and the Schmidt states $ \alpha_{2n-1}, \alpha_{2n+1}$ to the left and right, respectively. A singular value decomposition separating the physical indices $\sigma_{2n-1}, \sigma_{2n}$ allows a projection to a new single-site $\tilde{\sigma_n}$ and directly yields the corresponding isometry for the MERA layer.

Together these two steps produce a unitary sub-layer of $L'/2-1$ tensors and an isometry sub-layer of $L'/2$ tensors. We then repeat this process with our new MPS of length $L'/2$ to produce the next layer in the MERA network.

\begin{figure*}
    \centering
    \includegraphics[width=\textwidth]{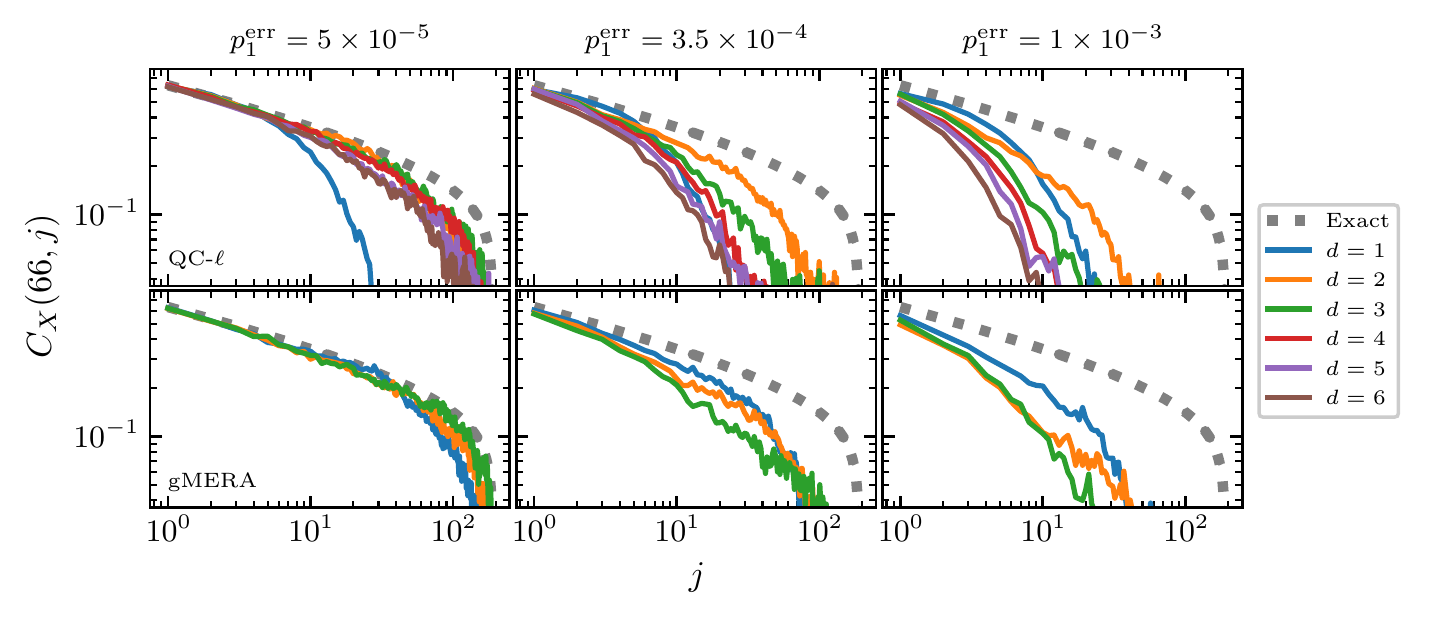}
    \caption{Simulations of ground states for the unperturbed TFI model realized by QC-$\ell$ (top row) and gMERA (bottom) networks for different probabilities of depolarizing noise. Two qubit error rates are chosen as $p^{\mathrm{err}}_{2} = 10 p^{\mathrm{err}}_{1}$. A depth $d$ QC-$\ell$ circuit requries $d+1$ qubits compared to the $2d+1$ needed for (g)MERA.}
    \label{fig:noise_MPS}
\end{figure*}

\begin{figure}
    \centering
    \includegraphics[width=0.47\textwidth]{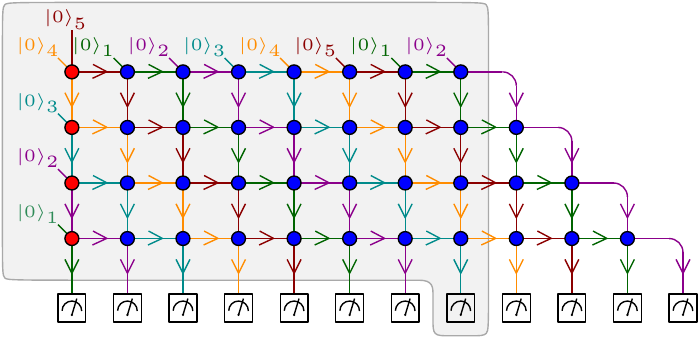}
    \caption{Isometric representation for unitary structure of depth 4 for a chain of 12 sites. Five qubits, which are color coded, are required for holographic simulation. The red and blue tensors are isometries and unitaries, respectively. The causal cone for site $8$ is shaded in grey, and for a depth $d$ circuit for a state of length $L$, there are $\mathcal{O}(dL)$ gates/tensors in the causal cone.}
    \label{fig:unitary}
\end{figure}

\subsection{Splitting of a gMERA layer}
To produce the gMERA layer by layer, we again start with an MPS of length $L'$ in right canonical form. Rather than generate local unitaries and isometries that act on a local pair of sites, we wish to generate "global" unitary and isometry matrix product operators (MPOs) that act on the entire MPS. To do this, we use the Moses Move algorithm developed in the context of 2D isometric tensor networks \cite{zaletel2020}. This algorithm splits a two-sided MPS $\ket{\Psi}$ into the product of an isometric MPO $A$ and a new MPS $\ket{\Phi}$, with $\ket{\Psi} \approx A \ket{\Phi}$. By specifying the bond dimensions of the virtual legs between $A$ and $\ket{\Phi}$, $A$ can be either a global unitary or an isometry, while the virtual bond dimension between tensors in $A$ controls on how many sites each tensor in $A$ can act. We set all bond dimensions of A, i.e. both vertical and horizontal bonds in Fig.~\ref{fig:isoMERA}, to $2$ so that the produced network can be directly viewed as a quantum circuit of two-site unitary gates. To produce a layer of a gMERA, first the MM algorithm is applied to our current MPS of length $L'$ to produce an almost unitary $A$ MPO and a new MPS of length $L'-1$. The $A$ is almost a unitary, but the last, leftmost tensor in $A_U$ is actually an isometry (projects from an incoming bond dimension of 4 to 2); see the bottom-most row of tensors in Fig.~\ref{fig:isoMERA}, where the leftmost tensor is an isometry. We then use the MM algorithm again to produce an isometric MPO $A_I$ and an MPS of length $L'/2$, where the tensors in $A_I$ alternate between unitaries and isometries. This MPO $A_I$ projects from $L'-1$ sites to $L'/2$ sites.

A gMERA layer is formed by stacking $A_I$ on top of $A_U$ as in Fig.~\ref{fig:isoMERA}. We repeat this procedure to produce the desired number of layers.

\subsection{Network Optimization}
Once the MERA and gMERA structures are produced, we optimize the networks by maximizing overlap with the original ground state MPS of $L$ sites found by DMRG. This is done in a manner similar to the Evenbly-Vidal algorithm of successive polar decompositions for minimizing the energy of a ternary MERA \cite{evenbly2009}. For the tensor to be optimized, we iteratively form its environment by contracting all tensors in the overlap except for the tensor of interest. We then replace this tensor with the isometric tensor resulting from the polar decomposition of the environment, where the dimensions of the environment determine whether this tensor is a unitary or a projector. Each cycle of this iterative process involves updating each tensor, and we do enough iterations so that the fidelity is no longer meaningfully increasing, e.g. typically 1000 cycles. This produces the optimized finite MERA and gMERA networks that we convert into quantum circuits as discussed in the main paper.

\subsection{Conversion to Quantum Circuit}
Once we have an optimized tensor network, we convert this into a quantum circuit by expanding each isometry into a two-qubit unitary.
As noted in the main text, an arbitrary $SU(2)$ gate can be converted into a sequence of 7 single-qubit unitaries specified by 15 angles and three CNOTS \cite{vatan2004}.
We find these angles by numerically minimizing (via \texttt{scipy.optimize}) the norm of the difference between the desired unitary gate and the unitary gate parameterized by the angles.

As noted, in the interior of the MERA, one can exploit the gauge freedom on contracted indices of the MERA to reduce this further to just 9 free angles per two-qubit unitary. 
This can be done as the last generic single qubit rotation from one gate and the first generic single qubit from the next gate connected by a bond can be combined; hence we can remove two generic single-qubit rotations from each gate, removing the need for 6 angles.
For models with charge conservation, the number of free parameters can be reduced even further by enforcing a block-diagonal form of the tensors, similar as in classical tensor network simulations \cite{singh2010,niu2021}.

\section{Experimental Details}
The Quantinuum's System Model H1 quantum computer is a trapped-ion system based on a quantum charge-coupled device (QCCD) architecture, where quantum information is stored in the hyperfine ``clock'' states of $^{171}\mathrm{Yb}^{+}$ ions that can be dynamically moved between different areas of the processor \cite{pino2021}.
Each qubit is a pair of a $^{171}\mathrm{Yb}^{+}$ qubit ion and a $^{138}\mathrm{Ba}^{+}$ ion used for sympathetic cooling.
Two qubits are paired together to form a four-ion ``crystal", with each ion crystal placed in a gate processing zone where single-qubit and entangling two-qubit gates can be applied to both qubits.
Ion crystals can be split, moved between gate zones, and merged by dynamical electrical fields created by a cryogenic surface trap.
Operations in different gate zones can be done in parallel.
The isolation of ion crystals in different zones of the processor lead to long coherence times.

For depth $d=1$ $(2)$, we run 3 (2) MERA networks in parallel to obtain the required measurement samples more efficiently.
This uses 9 (10) qubits and allows us to reduce quantum computing costs.
Further, for depth $d=1,2$ MERA circuits, we do not need to simulate the entire chain of length $L$, as inherently with the MERA ansatz only a finite range of sites to the right can have non-zero correlations with starting site $10$.
Hence we simulate up to site 17 for depth 1 MERA and site 25 for depth 2 MERA.
For depths 3 and 4, the entire chain is simulated as non-zero correlations are possible over the entire chain.

\section{Unitary networks}
\label{sec:ladder}
In the main text, we introduced the gMERA as a natural interpolation between MERA and MPS networks.
Here we provide results for a network ansatz that can be viewed as a simple quantum circuit analog of an MPS. 
This ansatz is composed entirely of the global unitary sub-layers used in the gMERA, as shown in Fig.~\ref{fig:unitary}. 
We label these as $QC_{\ell}$ networks, following the terminology of \cite{haghshenas2021}, as they can be viewed as quantum circuits formed by stacking ladder circuits, starting at the rightmost spin and ascending up to the left.
These unitary layers should be contrasted with the brick-wall pattern used in MERA.
A depth $d$ QC-$\ell$ network can be collapsed to a $\chi=2^d$ MPS, where the MPS tensor is formed by collapsing all the tensors in a column.

While the number of gates in a MERA and gMERA network for a system of size $L$ asymptotes to $2L$ and $4L$, respectively, as a function of depth $d$, the number of gates in the QC-$\ell$ network scales as $\mathcal{O}(dL)$. 
This is natural as these circuits do not have the coarse-graining properties of MERA and gMERA. 
As with gMERA, a single unitary layer is sufficient to give an infinite range of correlations, but again these correlations decay exponentially.
Additionally, the causal cone for a site is extensive and includes all tensors above and to the left, so like gMERA these networks are not suitable for top-down holographic simulation.

The QC-$\ell$ circuits are a generalization of the gMERA, as we replace all isometry tensors except for those in the leftmost column with unitaries and add non-trivial tensors to empty sites in the gMERA.
Thus the representational power can increases, at the expense of a gate count that scales linearly with depth.
Note that a $d=2$ QC-$\ell$ network and $d=1$ gMERA are equivalent to one another.
We perform noise analysis of these network in Fig.~\ref{fig:noise_MPS}, where we see that QC-$\ell$ networks are more strongly impacted by noise than gMERA; a depth $4$ QC-$\ell$ and depth $2$ gMERA use the same number of qubits, yet the performance of the QC-$\ell$ circuit degrades more significantly with increased noise rates.
The performance of higher depth networks decays quickly with noise, as with approximately $L$ tensors per layer, errors are equally likely in higher layers as in lower layers.
Errors in higher layers will affect all tensors to the right and below, even though errors will decay with the correlation length of the state. 
For MERA and gMERA networks, due to the coarse-graining, a layer has roughly half as many gates as the layer below it, implying that errors tend to occur lower in the network and thus affect fewer future gates.

\bibliographystyle{apsrev4-1}
\bibliography{main}

\begin{thebibliography}{40}%
\makeatletter
\providecommand \@ifxundefined [1]{%
 \@ifx{#1\undefined}
}%
\providecommand \@ifnum [1]{%
 \ifnum #1\expandafter \@firstoftwo
 \else \expandafter \@secondoftwo
 \fi
}%
\providecommand \@ifx [1]{%
 \ifx #1\expandafter \@firstoftwo
 \else \expandafter \@secondoftwo
 \fi
}%
\providecommand \natexlab [1]{#1}%
\providecommand \enquote  [1]{``#1''}%
\providecommand \bibnamefont  [1]{#1}%
\providecommand \bibfnamefont [1]{#1}%
\providecommand \citenamefont [1]{#1}%
\providecommand \href@noop [0]{\@secondoftwo}%
\providecommand \href [0]{\begingroup \@sanitize@url \@href}%
\providecommand \@href[1]{\@@startlink{#1}\@@href}%
\providecommand \@@href[1]{\endgroup#1\@@endlink}%
\providecommand \@sanitize@url [0]{\catcode `\\12\catcode `\$12\catcode
  `\&12\catcode `\#12\catcode `\^12\catcode `\_12\catcode `\%12\relax}%
\providecommand \@@startlink[1]{}%
\providecommand \@@endlink[0]{}%
\providecommand \url  [0]{\begingroup\@sanitize@url \@url }%
\providecommand \@url [1]{\endgroup\@href {#1}{\urlprefix }}%
\providecommand \urlprefix  [0]{URL }%
\providecommand \Eprint [0]{\href }%
\providecommand \doibase [0]{http://dx.doi.org/}%
\providecommand \selectlanguage [0]{\@gobble}%
\providecommand \bibinfo  [0]{\@secondoftwo}%
\providecommand \bibfield  [0]{\@secondoftwo}%
\providecommand \translation [1]{[#1]}%
\providecommand \BibitemOpen [0]{}%
\providecommand \bibitemStop [0]{}%
\providecommand \bibitemNoStop [0]{.\EOS\space}%
\providecommand \EOS [0]{\spacefactor3000\relax}%
\providecommand \BibitemShut  [1]{\csname bibitem#1\endcsname}%
\let\auto@bib@innerbib\@empty
\bibitem [{\citenamefont {Preskill}(2018)}]{preskill2018}%
  \BibitemOpen
  \bibfield  {author} {\bibinfo {author} {\bibfnamefont {J.}~\bibnamefont
  {Preskill}},\ }\href {\doibase 10.22331/q-2018-08-06-79} {\bibfield
  {journal} {\bibinfo  {journal} {Quantum}\ }\textbf {\bibinfo {volume} {2}},\
  \bibinfo {pages} {79} (\bibinfo {year} {2018})}\BibitemShut {NoStop}%
\bibitem [{\citenamefont {Sch{\"o}n}\ \emph {et~al.}(2005)\citenamefont
  {Sch{\"o}n}, \citenamefont {Solano}, \citenamefont {Verstraete},
  \citenamefont {Cirac},\ and\ \citenamefont {Wolf}}]{schon2005sequential}%
  \BibitemOpen
  \bibfield  {author} {\bibinfo {author} {\bibfnamefont {C.}~\bibnamefont
  {Sch{\"o}n}}, \bibinfo {author} {\bibfnamefont {E.}~\bibnamefont {Solano}},
  \bibinfo {author} {\bibfnamefont {F.}~\bibnamefont {Verstraete}}, \bibinfo
  {author} {\bibfnamefont {J.~I.}\ \bibnamefont {Cirac}}, \ and\ \bibinfo
  {author} {\bibfnamefont {M.~M.}\ \bibnamefont {Wolf}},\ }\href {\doibase
  10.1103/PhysRevLett.95.110503} {\bibfield  {journal} {\bibinfo  {journal}
  {Physical Review Letters}\ }\textbf {\bibinfo {volume} {95}},\ \bibinfo
  {pages} {110503} (\bibinfo {year} {2005})}\BibitemShut {NoStop}%
\bibitem [{\citenamefont {Kim}(2017{\natexlab{a}})}]{kim2017}%
  \BibitemOpen
  \bibfield  {author} {\bibinfo {author} {\bibfnamefont {I.~H.}\ \bibnamefont
  {Kim}},\ }\href@noop {} {\  (\bibinfo {year} {2017}{\natexlab{a}})},\ \Eprint
  {http://arxiv.org/abs/1702.02093} {arXiv:1702.02093 [quant-ph]} \BibitemShut
  {NoStop}%
\bibitem [{\citenamefont {Kim}(2017{\natexlab{b}})}]{kim2017b}%
  \BibitemOpen
  \bibfield  {author} {\bibinfo {author} {\bibfnamefont {I.~H.}\ \bibnamefont
  {Kim}},\ }\href@noop {} {\  (\bibinfo {year} {2017}{\natexlab{b}})},\ \Eprint
  {http://arxiv.org/abs/1703.00032} {arXiv:1703.00032 [quant-ph]} \BibitemShut
  {NoStop}%
\bibitem [{\citenamefont {Barratt}\ \emph {et~al.}(2021)\citenamefont
  {Barratt}, \citenamefont {Dborin}, \citenamefont {Bal}, \citenamefont
  {Stojevic}, \citenamefont {Pollmann},\ and\ \citenamefont
  {Green}}]{barratt2020parallel}%
  \BibitemOpen
  \bibfield  {author} {\bibinfo {author} {\bibfnamefont {F.}~\bibnamefont
  {Barratt}}, \bibinfo {author} {\bibfnamefont {J.}~\bibnamefont {Dborin}},
  \bibinfo {author} {\bibfnamefont {M.}~\bibnamefont {Bal}}, \bibinfo {author}
  {\bibfnamefont {V.}~\bibnamefont {Stojevic}}, \bibinfo {author}
  {\bibfnamefont {F.}~\bibnamefont {Pollmann}}, \ and\ \bibinfo {author}
  {\bibfnamefont {A.~G.}\ \bibnamefont {Green}},\ }\href {\doibase
  10.1038/s41534-021-00420-3} {\bibfield  {journal} {\bibinfo  {journal} {npj
  Quantum Information}\ }\textbf {\bibinfo {volume} {7}} (\bibinfo {year}
  {2021}),\ 10.1038/s41534-021-00420-3}\BibitemShut {NoStop}%
\bibitem [{\citenamefont {Foss-Feig}\ \emph
  {et~al.}(2021{\natexlab{a}})\citenamefont {Foss-Feig}, \citenamefont {Hayes},
  \citenamefont {Dreiling}, \citenamefont {Figgatt}, \citenamefont {Gaebler},
  \citenamefont {Moses}, \citenamefont {Pino},\ and\ \citenamefont
  {Potter}}]{Foss-Feig2021a}%
  \BibitemOpen
  \bibfield  {author} {\bibinfo {author} {\bibfnamefont {M.}~\bibnamefont
  {Foss-Feig}}, \bibinfo {author} {\bibfnamefont {D.}~\bibnamefont {Hayes}},
  \bibinfo {author} {\bibfnamefont {J.~M.}\ \bibnamefont {Dreiling}}, \bibinfo
  {author} {\bibfnamefont {C.}~\bibnamefont {Figgatt}}, \bibinfo {author}
  {\bibfnamefont {J.~P.}\ \bibnamefont {Gaebler}}, \bibinfo {author}
  {\bibfnamefont {S.~A.}\ \bibnamefont {Moses}}, \bibinfo {author}
  {\bibfnamefont {J.~M.}\ \bibnamefont {Pino}}, \ and\ \bibinfo {author}
  {\bibfnamefont {A.~C.}\ \bibnamefont {Potter}},\ }\href {\doibase
  10.1103/physrevresearch.3.033002} {\bibfield  {journal} {\bibinfo  {journal}
  {Physical Review Research}\ }\textbf {\bibinfo {volume} {3}} (\bibinfo {year}
  {2021}{\natexlab{a}}),\ 10.1103/physrevresearch.3.033002}\BibitemShut
  {NoStop}%
\bibitem [{\citenamefont {Östlund}\ and\ \citenamefont
  {Rommer}(1995)}]{ostlund1995}%
  \BibitemOpen
  \bibfield  {author} {\bibinfo {author} {\bibfnamefont {S.}~\bibnamefont
  {Östlund}}\ and\ \bibinfo {author} {\bibfnamefont {S.}~\bibnamefont
  {Rommer}},\ }\href {\doibase 10.1103/physrevlett.75.3537} {\bibfield
  {journal} {\bibinfo  {journal} {Physical Review Letters}\ }\textbf {\bibinfo
  {volume} {75}},\ \bibinfo {pages} {3537–3540} (\bibinfo {year}
  {1995})}\BibitemShut {NoStop}%
\bibitem [{\citenamefont {Verstraete}\ and\ \citenamefont
  {Cirac}(2006)}]{verstraete2006}%
  \BibitemOpen
  \bibfield  {author} {\bibinfo {author} {\bibfnamefont {F.}~\bibnamefont
  {Verstraete}}\ and\ \bibinfo {author} {\bibfnamefont {J.~I.}\ \bibnamefont
  {Cirac}},\ }\href {\doibase 10.1103/physrevb.73.094423} {\bibfield  {journal}
  {\bibinfo  {journal} {Physical Review B}\ }\textbf {\bibinfo {volume} {73}}
  (\bibinfo {year} {2006}),\ 10.1103/physrevb.73.094423}\BibitemShut {NoStop}%
\bibitem [{\citenamefont {Hastings}(2007)}]{hastings2007}%
  \BibitemOpen
  \bibfield  {author} {\bibinfo {author} {\bibfnamefont {M.~B.}\ \bibnamefont
  {Hastings}},\ }\href {\doibase 10.1088/1742-5468/2007/08/p08024} {\bibfield
  {journal} {\bibinfo  {journal} {Journal of Statistical Mechanics: Theory and
  Experiment}\ }\textbf {\bibinfo {volume} {2007}},\ \bibinfo {pages}
  {P08024–P08024} (\bibinfo {year} {2007})}\BibitemShut {NoStop}%
\bibitem [{\citenamefont {Zaletel}\ and\ \citenamefont
  {Pollmann}(2020)}]{zaletel2020}%
  \BibitemOpen
  \bibfield  {author} {\bibinfo {author} {\bibfnamefont {M.~P.}\ \bibnamefont
  {Zaletel}}\ and\ \bibinfo {author} {\bibfnamefont {F.}~\bibnamefont
  {Pollmann}},\ }\href {\doibase 10.1103/physrevlett.124.037201} {\bibfield
  {journal} {\bibinfo  {journal} {Physical Review Letters}\ }\textbf {\bibinfo
  {volume} {124}} (\bibinfo {year} {2020}),\
  10.1103/physrevlett.124.037201}\BibitemShut {NoStop}%
\bibitem [{\citenamefont {Haghshenas}\ \emph {et~al.}(2021)\citenamefont
  {Haghshenas}, \citenamefont {Gray}, \citenamefont {Potter},\ and\
  \citenamefont {Chan}}]{haghshenas2021}%
  \BibitemOpen
  \bibfield  {author} {\bibinfo {author} {\bibfnamefont {R.}~\bibnamefont
  {Haghshenas}}, \bibinfo {author} {\bibfnamefont {J.}~\bibnamefont {Gray}},
  \bibinfo {author} {\bibfnamefont {A.~C.}\ \bibnamefont {Potter}}, \ and\
  \bibinfo {author} {\bibfnamefont {G.~K.-L.}\ \bibnamefont {Chan}},\
  }\href@noop {} {\  (\bibinfo {year} {2021})},\ \Eprint
  {http://arxiv.org/abs/2107.01307} {arXiv:2107.01307 [quant-ph]} \BibitemShut
  {NoStop}%
\bibitem [{\citenamefont {Lin}\ \emph {et~al.}(2021)\citenamefont {Lin},
  \citenamefont {Dilip}, \citenamefont {Green}, \citenamefont {Smith},\ and\
  \citenamefont {Pollmann}}]{lin2021}%
  \BibitemOpen
  \bibfield  {author} {\bibinfo {author} {\bibfnamefont {S.-H.}\ \bibnamefont
  {Lin}}, \bibinfo {author} {\bibfnamefont {R.}~\bibnamefont {Dilip}}, \bibinfo
  {author} {\bibfnamefont {A.~G.}\ \bibnamefont {Green}}, \bibinfo {author}
  {\bibfnamefont {A.}~\bibnamefont {Smith}}, \ and\ \bibinfo {author}
  {\bibfnamefont {F.}~\bibnamefont {Pollmann}},\ }\href {\doibase
  10.1103/PRXQuantum.2.010342} {\bibfield  {journal} {\bibinfo  {journal} {PRX
  Quantum}\ }\textbf {\bibinfo {volume} {2}},\ \bibinfo {pages} {010342}
  (\bibinfo {year} {2021})}\BibitemShut {NoStop}%
\bibitem [{\citenamefont {Slattery}\ and\ \citenamefont
  {Clark}(2021)}]{slattery2021}%
  \BibitemOpen
  \bibfield  {author} {\bibinfo {author} {\bibfnamefont {L.}~\bibnamefont
  {Slattery}}\ and\ \bibinfo {author} {\bibfnamefont {B.~K.}\ \bibnamefont
  {Clark}},\ }\href@noop {} {\  (\bibinfo {year} {2021})},\ \Eprint
  {http://arxiv.org/abs/2108.02792} {arXiv:2108.02792 [quant-ph]} \BibitemShut
  {NoStop}%
\bibitem [{\citenamefont {Wei}\ \emph {et~al.}(2022{\natexlab{a}})\citenamefont
  {Wei}, \citenamefont {Malz},\ and\ \citenamefont {Cirac}}]{wei2022}%
  \BibitemOpen
  \bibfield  {author} {\bibinfo {author} {\bibfnamefont {Z.-Y.}\ \bibnamefont
  {Wei}}, \bibinfo {author} {\bibfnamefont {D.}~\bibnamefont {Malz}}, \ and\
  \bibinfo {author} {\bibfnamefont {J.~I.}\ \bibnamefont {Cirac}},\ }\href
  {\doibase 10.1103/physrevlett.128.010607} {\bibfield  {journal} {\bibinfo
  {journal} {Physical Review Letters}\ }\textbf {\bibinfo {volume} {128}}
  (\bibinfo {year} {2022}{\natexlab{a}}),\
  10.1103/physrevlett.128.010607}\BibitemShut {NoStop}%
\bibitem [{\citenamefont {Niu}\ \emph {et~al.}(2021)\citenamefont {Niu},
  \citenamefont {Haghshenas}, \citenamefont {Zhang}, \citenamefont {Foss-Feig},
  \citenamefont {Chan},\ and\ \citenamefont {Potter}}]{niu2021}%
  \BibitemOpen
  \bibfield  {author} {\bibinfo {author} {\bibfnamefont {D.}~\bibnamefont
  {Niu}}, \bibinfo {author} {\bibfnamefont {R.}~\bibnamefont {Haghshenas}},
  \bibinfo {author} {\bibfnamefont {Y.}~\bibnamefont {Zhang}}, \bibinfo
  {author} {\bibfnamefont {M.}~\bibnamefont {Foss-Feig}}, \bibinfo {author}
  {\bibfnamefont {G.~K.-L.}\ \bibnamefont {Chan}}, \ and\ \bibinfo {author}
  {\bibfnamefont {A.~C.}\ \bibnamefont {Potter}},\ }\href@noop {} {\  (\bibinfo
  {year} {2021})},\ \Eprint {http://arxiv.org/abs/2112.10810} {arXiv:2112.10810
  [cond-mat.str-el]} \BibitemShut {NoStop}%
\bibitem [{\citenamefont {Chertkov}\ \emph {et~al.}(2021)\citenamefont
  {Chertkov}, \citenamefont {Bohnet}, \citenamefont {Francois}, \citenamefont
  {Gaebler}, \citenamefont {Gresh}, \citenamefont {Hankin}, \citenamefont
  {Lee}, \citenamefont {Tobey}, \citenamefont {Hayes}, \citenamefont
  {Neyenhuis}, \citenamefont {Stutz}, \citenamefont {Potter},\ and\
  \citenamefont {Foss-Feig}}]{chertkov2021}%
  \BibitemOpen
  \bibfield  {author} {\bibinfo {author} {\bibfnamefont {E.}~\bibnamefont
  {Chertkov}}, \bibinfo {author} {\bibfnamefont {J.}~\bibnamefont {Bohnet}},
  \bibinfo {author} {\bibfnamefont {D.}~\bibnamefont {Francois}}, \bibinfo
  {author} {\bibfnamefont {J.}~\bibnamefont {Gaebler}}, \bibinfo {author}
  {\bibfnamefont {D.}~\bibnamefont {Gresh}}, \bibinfo {author} {\bibfnamefont
  {A.}~\bibnamefont {Hankin}}, \bibinfo {author} {\bibfnamefont
  {K.}~\bibnamefont {Lee}}, \bibinfo {author} {\bibfnamefont {R.}~\bibnamefont
  {Tobey}}, \bibinfo {author} {\bibfnamefont {D.}~\bibnamefont {Hayes}},
  \bibinfo {author} {\bibfnamefont {B.}~\bibnamefont {Neyenhuis}}, \bibinfo
  {author} {\bibfnamefont {R.}~\bibnamefont {Stutz}}, \bibinfo {author}
  {\bibfnamefont {A.~C.}\ \bibnamefont {Potter}}, \ and\ \bibinfo {author}
  {\bibfnamefont {M.}~\bibnamefont {Foss-Feig}},\ }\href@noop {} {\  (\bibinfo
  {year} {2021})},\ \Eprint {http://arxiv.org/abs/2105.09324} {arXiv:2105.09324
  [quant-ph]} \BibitemShut {NoStop}%
\bibitem [{\citenamefont {Foss-Feig}\ \emph
  {et~al.}(2021{\natexlab{b}})\citenamefont {Foss-Feig}, \citenamefont
  {Ragole}, \citenamefont {Potter}, \citenamefont {Dreiling}, \citenamefont
  {Figgatt}, \citenamefont {Gaebler}, \citenamefont {Hall}, \citenamefont
  {Moses}, \citenamefont {Pino}, \citenamefont {Spaun}, \citenamefont
  {Neyenhuis},\ and\ \citenamefont {Hayes}}]{Foss-Feig2021b}%
  \BibitemOpen
  \bibfield  {author} {\bibinfo {author} {\bibfnamefont {M.}~\bibnamefont
  {Foss-Feig}}, \bibinfo {author} {\bibfnamefont {S.}~\bibnamefont {Ragole}},
  \bibinfo {author} {\bibfnamefont {A.}~\bibnamefont {Potter}}, \bibinfo
  {author} {\bibfnamefont {J.}~\bibnamefont {Dreiling}}, \bibinfo {author}
  {\bibfnamefont {C.}~\bibnamefont {Figgatt}}, \bibinfo {author} {\bibfnamefont
  {J.}~\bibnamefont {Gaebler}}, \bibinfo {author} {\bibfnamefont
  {A.}~\bibnamefont {Hall}}, \bibinfo {author} {\bibfnamefont {S.}~\bibnamefont
  {Moses}}, \bibinfo {author} {\bibfnamefont {J.}~\bibnamefont {Pino}},
  \bibinfo {author} {\bibfnamefont {B.}~\bibnamefont {Spaun}}, \bibinfo
  {author} {\bibfnamefont {B.}~\bibnamefont {Neyenhuis}}, \ and\ \bibinfo
  {author} {\bibfnamefont {D.}~\bibnamefont {Hayes}},\ }\href@noop {} {\
  (\bibinfo {year} {2021}{\natexlab{b}})},\ \Eprint
  {http://arxiv.org/abs/2104.11235} {arXiv:2104.11235 [quant-ph]} \BibitemShut
  {NoStop}%
\bibitem [{\citenamefont {Vidal}(2007)}]{vidal2007}%
  \BibitemOpen
  \bibfield  {author} {\bibinfo {author} {\bibfnamefont {G.}~\bibnamefont
  {Vidal}},\ }\href {\doibase 10.1103/physrevlett.99.220405} {\bibfield
  {journal} {\bibinfo  {journal} {Physical Review Letters}\ }\textbf {\bibinfo
  {volume} {99}} (\bibinfo {year} {2007}),\
  10.1103/physrevlett.99.220405}\BibitemShut {NoStop}%
\bibitem [{\citenamefont {Vidal}(2008)}]{vidal2008}%
  \BibitemOpen
  \bibfield  {author} {\bibinfo {author} {\bibfnamefont {G.}~\bibnamefont
  {Vidal}},\ }\href {\doibase 10.1103/physrevlett.101.110501} {\bibfield
  {journal} {\bibinfo  {journal} {Physical Review Letters}\ }\textbf {\bibinfo
  {volume} {101}} (\bibinfo {year} {2008}),\
  10.1103/physrevlett.101.110501}\BibitemShut {NoStop}%
\bibitem [{\citenamefont {Evenbly}\ and\ \citenamefont
  {Vidal}(2013)}]{evenbly2013}%
  \BibitemOpen
  \bibfield  {author} {\bibinfo {author} {\bibfnamefont {G.}~\bibnamefont
  {Evenbly}}\ and\ \bibinfo {author} {\bibfnamefont {G.}~\bibnamefont
  {Vidal}},\ }\href@noop {} {\  (\bibinfo {year} {2013})},\ \Eprint
  {http://arxiv.org/abs/1109.5334} {arXiv:1109.5334 [quant-ph]} \BibitemShut
  {NoStop}%
\bibitem [{\citenamefont {Evenbly}\ and\ \citenamefont
  {Vidal}(2014)}]{evenbly2014}%
  \BibitemOpen
  \bibfield  {author} {\bibinfo {author} {\bibfnamefont {G.}~\bibnamefont
  {Evenbly}}\ and\ \bibinfo {author} {\bibfnamefont {G.}~\bibnamefont
  {Vidal}},\ }\href {\doibase 10.1103/physrevb.89.235113} {\bibfield  {journal}
  {\bibinfo  {journal} {Physical Review B}\ }\textbf {\bibinfo {volume} {89}}
  (\bibinfo {year} {2014}),\ 10.1103/physrevb.89.235113}\BibitemShut {NoStop}%
\bibitem [{\citenamefont {Pino}\ \emph {et~al.}(2021)\citenamefont {Pino},
  \citenamefont {Dreiling}, \citenamefont {Figgatt}, \citenamefont {Gaebler},
  \citenamefont {Moses}, \citenamefont {Allman}, \citenamefont {Baldwin},
  \citenamefont {Foss-Feig}, \citenamefont {Hayes}, \citenamefont {Mayer},\
  and\ \citenamefont {et~al.}}]{pino2021}%
  \BibitemOpen
  \bibfield  {author} {\bibinfo {author} {\bibfnamefont {J.~M.}\ \bibnamefont
  {Pino}}, \bibinfo {author} {\bibfnamefont {J.~M.}\ \bibnamefont {Dreiling}},
  \bibinfo {author} {\bibfnamefont {C.}~\bibnamefont {Figgatt}}, \bibinfo
  {author} {\bibfnamefont {J.~P.}\ \bibnamefont {Gaebler}}, \bibinfo {author}
  {\bibfnamefont {S.~A.}\ \bibnamefont {Moses}}, \bibinfo {author}
  {\bibfnamefont {M.~S.}\ \bibnamefont {Allman}}, \bibinfo {author}
  {\bibfnamefont {C.~H.}\ \bibnamefont {Baldwin}}, \bibinfo {author}
  {\bibfnamefont {M.}~\bibnamefont {Foss-Feig}}, \bibinfo {author}
  {\bibfnamefont {D.}~\bibnamefont {Hayes}}, \bibinfo {author} {\bibfnamefont
  {K.}~\bibnamefont {Mayer}}, \ and\ \bibinfo {author} {\bibnamefont
  {et~al.}},\ }\href {\doibase 10.1038/s41586-021-03318-4} {\bibfield
  {journal} {\bibinfo  {journal} {Nature}\ }\textbf {\bibinfo {volume} {592}},\
  \bibinfo {pages} {209–213} (\bibinfo {year} {2021})}\BibitemShut {NoStop}%
\bibitem [{\citenamefont {Kim}\ and\ \citenamefont {Swingle}(2017)}]{kim2017c}%
  \BibitemOpen
  \bibfield  {author} {\bibinfo {author} {\bibfnamefont {I.~H.}\ \bibnamefont
  {Kim}}\ and\ \bibinfo {author} {\bibfnamefont {B.}~\bibnamefont {Swingle}},\
  }\href@noop {} {\  (\bibinfo {year} {2017})},\ \Eprint
  {http://arxiv.org/abs/1711.07500} {arXiv:1711.07500 [quant-ph]} \BibitemShut
  {NoStop}%
\bibitem [{\citenamefont {Sewell}\ and\ \citenamefont
  {Jordan}(2021)}]{sewell2021}%
  \BibitemOpen
  \bibfield  {author} {\bibinfo {author} {\bibfnamefont {T.~J.}\ \bibnamefont
  {Sewell}}\ and\ \bibinfo {author} {\bibfnamefont {S.~P.}\ \bibnamefont
  {Jordan}},\ }\href@noop {} {\  (\bibinfo {year} {2021})},\ \Eprint
  {http://arxiv.org/abs/2109.09787} {arXiv:2109.09787 [quant-ph]} \BibitemShut
  {NoStop}%
\bibitem [{\citenamefont {Evenbly}\ and\ \citenamefont
  {Vidal}(2009)}]{evenbly2009}%
  \BibitemOpen
  \bibfield  {author} {\bibinfo {author} {\bibfnamefont {G.}~\bibnamefont
  {Evenbly}}\ and\ \bibinfo {author} {\bibfnamefont {G.}~\bibnamefont
  {Vidal}},\ }\href {\doibase 10.1103/physrevb.79.144108} {\bibfield  {journal}
  {\bibinfo  {journal} {Physical Review B}\ }\textbf {\bibinfo {volume} {79}}
  (\bibinfo {year} {2009}),\ 10.1103/physrevb.79.144108}\BibitemShut {NoStop}%
\bibitem [{\citenamefont {Vatan}\ and\ \citenamefont
  {Williams}(2004)}]{vatan2004}%
  \BibitemOpen
  \bibfield  {author} {\bibinfo {author} {\bibfnamefont {F.}~\bibnamefont
  {Vatan}}\ and\ \bibinfo {author} {\bibfnamefont {C.}~\bibnamefont
  {Williams}},\ }\href {\doibase 10.1103/physreva.69.032315} {\bibfield
  {journal} {\bibinfo  {journal} {Physical Review A}\ }\textbf {\bibinfo
  {volume} {69}} (\bibinfo {year} {2004}),\
  10.1103/physreva.69.032315}\BibitemShut {NoStop}%
\bibitem [{\citenamefont {Singh}\ \emph {et~al.}(2010)\citenamefont {Singh},
  \citenamefont {Pfeifer},\ and\ \citenamefont {Vidal}}]{singh2010}%
  \BibitemOpen
  \bibfield  {author} {\bibinfo {author} {\bibfnamefont {S.}~\bibnamefont
  {Singh}}, \bibinfo {author} {\bibfnamefont {R.~N.~C.}\ \bibnamefont
  {Pfeifer}}, \ and\ \bibinfo {author} {\bibfnamefont {G.}~\bibnamefont
  {Vidal}},\ }\href {\doibase 10.1103/PhysRevA.82.050301} {\bibfield  {journal}
  {\bibinfo  {journal} {Phys. Rev. A}\ }\textbf {\bibinfo {volume} {82}},\
  \bibinfo {pages} {050301} (\bibinfo {year} {2010})}\BibitemShut {NoStop}%
\bibitem [{\citenamefont {Moll}\ \emph {et~al.}(2018)\citenamefont {Moll},
  \citenamefont {Barkoutsos}, \citenamefont {Bishop}, \citenamefont {Chow},
  \citenamefont {Cross}, \citenamefont {Egger}, \citenamefont {Filipp},
  \citenamefont {Fuhrer}, \citenamefont {Gambetta}, \citenamefont {Ganzhorn},
  \citenamefont {Kandala}, \citenamefont {Mezzacapo}, \citenamefont {Müller},
  \citenamefont {Riess}, \citenamefont {Salis}, \citenamefont {Smolin},
  \citenamefont {Tavernelli},\ and\ \citenamefont {Temme}}]{Moll2018}%
  \BibitemOpen
  \bibfield  {author} {\bibinfo {author} {\bibfnamefont {N.}~\bibnamefont
  {Moll}}, \bibinfo {author} {\bibfnamefont {P.}~\bibnamefont {Barkoutsos}},
  \bibinfo {author} {\bibfnamefont {L.~S.}\ \bibnamefont {Bishop}}, \bibinfo
  {author} {\bibfnamefont {J.~M.}\ \bibnamefont {Chow}}, \bibinfo {author}
  {\bibfnamefont {A.}~\bibnamefont {Cross}}, \bibinfo {author} {\bibfnamefont
  {D.~J.}\ \bibnamefont {Egger}}, \bibinfo {author} {\bibfnamefont
  {S.}~\bibnamefont {Filipp}}, \bibinfo {author} {\bibfnamefont
  {A.}~\bibnamefont {Fuhrer}}, \bibinfo {author} {\bibfnamefont {J.~M.}\
  \bibnamefont {Gambetta}}, \bibinfo {author} {\bibfnamefont {M.}~\bibnamefont
  {Ganzhorn}}, \bibinfo {author} {\bibfnamefont {A.}~\bibnamefont {Kandala}},
  \bibinfo {author} {\bibfnamefont {A.}~\bibnamefont {Mezzacapo}}, \bibinfo
  {author} {\bibfnamefont {P.}~\bibnamefont {Müller}}, \bibinfo {author}
  {\bibfnamefont {W.}~\bibnamefont {Riess}}, \bibinfo {author} {\bibfnamefont
  {G.}~\bibnamefont {Salis}}, \bibinfo {author} {\bibfnamefont
  {J.}~\bibnamefont {Smolin}}, \bibinfo {author} {\bibfnamefont
  {I.}~\bibnamefont {Tavernelli}}, \ and\ \bibinfo {author} {\bibfnamefont
  {K.}~\bibnamefont {Temme}},\ }\href {\doibase 10.1088/2058-9565/aab822}
  {\bibfield  {journal} {\bibinfo  {journal} {Quantum Science and Technology}\
  }\textbf {\bibinfo {volume} {3}},\ \bibinfo {pages} {030503} (\bibinfo {year}
  {2018})}\BibitemShut {NoStop}%
\bibitem [{\citenamefont {Bal}\ \emph {et~al.}(2016)\citenamefont {Bal},
  \citenamefont {Rams}, \citenamefont {Zauner}, \citenamefont {Haegeman},\ and\
  \citenamefont {Verstraete}}]{bal2016}%
  \BibitemOpen
  \bibfield  {author} {\bibinfo {author} {\bibfnamefont {M.}~\bibnamefont
  {Bal}}, \bibinfo {author} {\bibfnamefont {M.~M.}\ \bibnamefont {Rams}},
  \bibinfo {author} {\bibfnamefont {V.}~\bibnamefont {Zauner}}, \bibinfo
  {author} {\bibfnamefont {J.}~\bibnamefont {Haegeman}}, \ and\ \bibinfo
  {author} {\bibfnamefont {F.}~\bibnamefont {Verstraete}},\ }\href {\doibase
  10.1103/physrevb.94.205122} {\bibfield  {journal} {\bibinfo  {journal}
  {Physical Review B}\ }\textbf {\bibinfo {volume} {94}} (\bibinfo {year}
  {2016}),\ 10.1103/physrevb.94.205122}\BibitemShut {NoStop}%
\bibitem [{\citenamefont {Radicevic}(2019)}]{radicevic2019spin}%
  \BibitemOpen
  \bibfield  {author} {\bibinfo {author} {\bibfnamefont {D.}~\bibnamefont
  {Radicevic}},\ }\href@noop {} {\enquote {\bibinfo {title} {Spin structures
  and exact dualities in low dimensions},}\ } (\bibinfo {year} {2019}),\
  \Eprint {http://arxiv.org/abs/1809.07757} {arXiv:1809.07757 [hep-th]}
  \BibitemShut {NoStop}%
\bibitem [{\citenamefont {Rahmani}\ \emph {et~al.}(2015)\citenamefont
  {Rahmani}, \citenamefont {Zhu}, \citenamefont {Franz},\ and\ \citenamefont
  {Affleck}}]{rahmani2015}%
  \BibitemOpen
  \bibfield  {author} {\bibinfo {author} {\bibfnamefont {A.}~\bibnamefont
  {Rahmani}}, \bibinfo {author} {\bibfnamefont {X.}~\bibnamefont {Zhu}},
  \bibinfo {author} {\bibfnamefont {M.}~\bibnamefont {Franz}}, \ and\ \bibinfo
  {author} {\bibfnamefont {I.}~\bibnamefont {Affleck}},\ }\href {\doibase
  10.1103/physrevb.92.235123} {\bibfield  {journal} {\bibinfo  {journal}
  {Physical Review B}\ }\textbf {\bibinfo {volume} {92}},\ \bibinfo {pages}
  {235123} (\bibinfo {year} {2015})}\BibitemShut {NoStop}%
\bibitem [{\citenamefont {Abraham}\ and\ \citenamefont {et~al}(2019)}]{Qiskit}%
  \BibitemOpen
  \bibfield  {author} {\bibinfo {author} {\bibfnamefont {H.}~\bibnamefont
  {Abraham}}\ and\ \bibinfo {author} {\bibnamefont {et~al}},\ }\href {\doibase
  10.5281/zenodo.2562110} {\  (\bibinfo {year} {2019}),\
  10.5281/zenodo.2562110}\BibitemShut {NoStop}%
\bibitem [{\citenamefont {Soejima}\ \emph {et~al.}(2020)\citenamefont
  {Soejima}, \citenamefont {Siva}, \citenamefont {Bultinck}, \citenamefont
  {Chatterjee}, \citenamefont {Pollmann},\ and\ \citenamefont
  {Zaletel}}]{soejima2020}%
  \BibitemOpen
  \bibfield  {author} {\bibinfo {author} {\bibfnamefont {T.}~\bibnamefont
  {Soejima}}, \bibinfo {author} {\bibfnamefont {K.}~\bibnamefont {Siva}},
  \bibinfo {author} {\bibfnamefont {N.}~\bibnamefont {Bultinck}}, \bibinfo
  {author} {\bibfnamefont {S.}~\bibnamefont {Chatterjee}}, \bibinfo {author}
  {\bibfnamefont {F.}~\bibnamefont {Pollmann}}, \ and\ \bibinfo {author}
  {\bibfnamefont {M.~P.}\ \bibnamefont {Zaletel}},\ }\href {\doibase
  10.1103/physrevb.101.085117} {\bibfield  {journal} {\bibinfo  {journal}
  {Physical Review B}\ }\textbf {\bibinfo {volume} {101}} (\bibinfo {year}
  {2020}),\ 10.1103/physrevb.101.085117}\BibitemShut {NoStop}%
\bibitem [{\citenamefont {Wei}\ \emph {et~al.}(2022{\natexlab{b}})\citenamefont
  {Wei}, \citenamefont {Malz},\ and\ \citenamefont
  {Cirac}}]{wei2022sequential}%
  \BibitemOpen
  \bibfield  {author} {\bibinfo {author} {\bibfnamefont {Z.-Y.}\ \bibnamefont
  {Wei}}, \bibinfo {author} {\bibfnamefont {D.}~\bibnamefont {Malz}}, \ and\
  \bibinfo {author} {\bibfnamefont {J.~I.}\ \bibnamefont {Cirac}},\ }\href
  {\doibase 10.1103/physrevlett.128.010607} {\bibfield  {journal} {\bibinfo
  {journal} {Physical Review Letters}\ }\textbf {\bibinfo {volume} {128}}
  (\bibinfo {year} {2022}{\natexlab{b}}),\
  10.1103/physrevlett.128.010607}\BibitemShut {NoStop}%
\bibitem [{\citenamefont {Schuch}\ \emph {et~al.}(2007)\citenamefont {Schuch},
  \citenamefont {Wolf}, \citenamefont {Verstraete},\ and\ \citenamefont
  {Cirac}}]{schuch2007}%
  \BibitemOpen
  \bibfield  {author} {\bibinfo {author} {\bibfnamefont {N.}~\bibnamefont
  {Schuch}}, \bibinfo {author} {\bibfnamefont {M.~M.}\ \bibnamefont {Wolf}},
  \bibinfo {author} {\bibfnamefont {F.}~\bibnamefont {Verstraete}}, \ and\
  \bibinfo {author} {\bibfnamefont {J.~I.}\ \bibnamefont {Cirac}},\ }\href
  {\doibase 10.1103/physrevlett.98.140506} {\bibfield  {journal} {\bibinfo
  {journal} {Physical Review Letters}\ }\textbf {\bibinfo {volume} {98}}
  (\bibinfo {year} {2007}),\ 10.1103/physrevlett.98.140506}\BibitemShut
  {NoStop}%
\bibitem [{\citenamefont {Peruzzo}\ \emph {et~al.}()\citenamefont {Peruzzo},
  \citenamefont {McClean}, \citenamefont {Shadbolt}, \citenamefont {Yung},
  \citenamefont {Zhou}, \citenamefont {Love}, \citenamefont {Aspuru-Guzik},\
  and\ \citenamefont {O'Brien}}]{peruzzo2014}%
  \BibitemOpen
  \bibfield  {author} {\bibinfo {author} {\bibfnamefont {A.}~\bibnamefont
  {Peruzzo}}, \bibinfo {author} {\bibfnamefont {J.}~\bibnamefont {McClean}},
  \bibinfo {author} {\bibfnamefont {P.}~\bibnamefont {Shadbolt}}, \bibinfo
  {author} {\bibfnamefont {M.-H.}\ \bibnamefont {Yung}}, \bibinfo {author}
  {\bibfnamefont {X.-Q.}\ \bibnamefont {Zhou}}, \bibinfo {author}
  {\bibfnamefont {P.~J.}\ \bibnamefont {Love}}, \bibinfo {author}
  {\bibfnamefont {A.}~\bibnamefont {Aspuru-Guzik}}, \ and\ \bibinfo {author}
  {\bibfnamefont {J.~L.}\ \bibnamefont {O'Brien}},\ }\href {\doibase
  10.1038/ncomms5213} {\bibfield  {journal} {\bibinfo  {journal} {Nature
  Communications}\ }\textbf {\bibinfo {volume} {5}},\ \bibinfo {pages}
  {4213}}\BibitemShut {NoStop}%
\bibitem [{\citenamefont {Vieijra}\ \emph {et~al.}(2021)\citenamefont
  {Vieijra}, \citenamefont {Haegeman}, \citenamefont {Verstraete},\ and\
  \citenamefont {Vanderstraeten}}]{vieijra2021}%
  \BibitemOpen
  \bibfield  {author} {\bibinfo {author} {\bibfnamefont {T.}~\bibnamefont
  {Vieijra}}, \bibinfo {author} {\bibfnamefont {J.}~\bibnamefont {Haegeman}},
  \bibinfo {author} {\bibfnamefont {F.}~\bibnamefont {Verstraete}}, \ and\
  \bibinfo {author} {\bibfnamefont {L.}~\bibnamefont {Vanderstraeten}},\ }\href
  {\doibase 10.1103/physrevb.104.235141} {\bibfield  {journal} {\bibinfo
  {journal} {Physical Review B}\ }\textbf {\bibinfo {volume} {104}} (\bibinfo
  {year} {2021}),\ 10.1103/physrevb.104.235141}\BibitemShut {NoStop}%
\bibitem [{\citenamefont {White}(1992)}]{white1992}%
  \BibitemOpen
  \bibfield  {author} {\bibinfo {author} {\bibfnamefont {S.~R.}\ \bibnamefont
  {White}},\ }\href {\doibase 10.1103/PhysRevLett.69.2863} {\bibfield
  {journal} {\bibinfo  {journal} {Phys. Rev. Lett.}\ }\textbf {\bibinfo
  {volume} {69}},\ \bibinfo {pages} {2863} (\bibinfo {year}
  {1992})}\BibitemShut {NoStop}%
\bibitem [{\citenamefont {Hauschild}\ and\ \citenamefont
  {Pollmann}(2018)}]{hauschild2018}%
  \BibitemOpen
  \bibfield  {author} {\bibinfo {author} {\bibfnamefont {J.}~\bibnamefont
  {Hauschild}}\ and\ \bibinfo {author} {\bibfnamefont {F.}~\bibnamefont
  {Pollmann}},\ }\href {\doibase 10.21468/scipostphyslectnotes.5} {\bibfield
  {journal} {\bibinfo  {journal} {SciPost Physics Lecture Notes}\ } (\bibinfo
  {year} {2018}),\ 10.21468/scipostphyslectnotes.5}\BibitemShut {NoStop}%
\bibitem [{\citenamefont {Hauschild}\ \emph {et~al.}(2018)\citenamefont
  {Hauschild}, \citenamefont {Leviatan}, \citenamefont {Bardarson},
  \citenamefont {Altman}, \citenamefont {Zaletel},\ and\ \citenamefont
  {Pollmann}}]{Hauschild2018b}%
  \BibitemOpen
  \bibfield  {author} {\bibinfo {author} {\bibfnamefont {J.}~\bibnamefont
  {Hauschild}}, \bibinfo {author} {\bibfnamefont {E.}~\bibnamefont {Leviatan}},
  \bibinfo {author} {\bibfnamefont {J.~H.}\ \bibnamefont {Bardarson}}, \bibinfo
  {author} {\bibfnamefont {E.}~\bibnamefont {Altman}}, \bibinfo {author}
  {\bibfnamefont {M.~P.}\ \bibnamefont {Zaletel}}, \ and\ \bibinfo {author}
  {\bibfnamefont {F.}~\bibnamefont {Pollmann}},\ }\href {\doibase
  10.1103/physrevb.98.235163} {\bibfield  {journal} {\bibinfo  {journal}
  {Physical Review B}\ }\textbf {\bibinfo {volume} {98}} (\bibinfo {year}
  {2018}),\ 10.1103/physrevb.98.235163}\BibitemShut {NoStop}%
\end{thebibliography}%
\end{document}